# Cooperative and non-linear phenomena at the neutral-ionic phase transition

Anna Painelli[a,*], Luca Del Freo[a] and Zoltan G. Soos[b]

[a] Dip. Chimica GIAF, Parma University & INSTM-UdR Parma, 43100 Parma, Italy

[b] Dip. of Chemistry, Princeton University, NJ-08544 Princeton, US

[*] Tel: +39 0521 905461; Fax: +39 0521 905556; E-mail: anna.painelli@unipr.it

___

**Abstract**

The complex interplay among on-site energy, Hubbard $U$, and coupling to Holstein and Peierls phonons at the neutral-ionic phase transition is discussed using diagrammatic valence bond calculations. The charge transfer and dimerization amplitudes and the infrared intensity of molecular vibrations are studied in the transition region.

___

The neutral-ionic phase transition (NIT), observed in charge-transfer (CT) crystals with a mixed donor-acceptor (DA) stack motif [1], is an interesting phenomenon driven by a complex interplay of charge, spin and vibrational degrees of freedom. The basic electronic model for a single DA stack is the standard Hubbard ($t,U$) model modified to have alternating site energies $\Delta(-1)^n$. With no loss of generality we set $2\Delta>0$ for the energy difference between A and D sites. For $\Delta>>U/2$, CT from D to A sites is suppressed and the mean ionicity, $\rho = 1+ <\Sigma_i (-1)^i n_i > / N$, approaches 0. In the opposite limit, $\Delta<<U/2$, CT is almost complete with $\rho\sim 1$. Varying $\Delta$ or $U$ changes $\rho$ smoothly between the two limits. Yet a critical line near

Δ-$U/2$ can be drawn in the $U,\Delta$ plane, separating a neutral (N) and an ionic (I) phase [2]. The two phases are distinguished based on the different nature of their ground state (GS) and excitation spectrum. The N phase is a band insulator with finite charge, spin and optical excitation gaps. The ionic phase is a Mott insulator, with a delocalized GS, finite charge and optical gaps, and vanishing magnetic (singlet-singlet and singlet-triplet) gaps. At the interface a metallic phase is established, as demonstrated by the simultaneous closing of charge and spin gaps, and by the finite charge-stiffness [2]. The metallic state at NIT is fairly exotic: it arises from the competition between $\Delta$ and $U$ that, taken individually, leads to band and Mott insulators, respectively.

In CT salts both lattice phonons and molecular vibrations play an active role via Peierls and Holstein coupling, respectively [3]. We include a single $k=0$ Peierls mode and a single $k=0$ Holstein mode, the extension to the multimode case being trivial. The Hamiltonian is ($t=1$):

$$H = -\sum_{i,s}\left[1+(-1)^i\sqrt{\frac{\varepsilon_d\omega_l}{2N}}(a^++a)\right]\left(c^+_{i,s}c_{i+1,s}+c^+_{i+1,s}c_{i,s}\right)$$
$$+\sum_{i,s}(-1)^i\left[\Delta+\sqrt{\frac{\varepsilon_{sp}\omega_v}{2N}}(b^++b)\right]c^+_{i,s}c_{i,s}+U\sum_i n_{i,s}n_{i,s'}+\omega_l\left(a^+a+\frac{1}{2}\right)+\omega_v\left(b^+b+\frac{1}{2}\right)$$

(1)

where $c_{i,s}$ is the annihilation operator for an electron with spin $s$ on site $i$, $a$ ($b$) is the annihilation operator for the Peierls (Holstein) phonon with frequency $\omega_l$ ($\omega_v$), the strength of the coupling being measured by $\varepsilon_d$ ($\varepsilon_{sp}$). The metallic state is unconditionally unstable: even in the $\varepsilon_d\to 0$ limit, the lattice phonon condenses to a finite dimerization, $\delta=\sqrt{\varepsilon_d\omega_l/2N}<a^++a>\neq 0$. This instability is clearly related to the Peierls instability of 1D metals. Unconditional dimerization is also recognized in the I regime, and is related to the vanishing of magnetic gaps. Dimerization energy is maximized at NIT, where both charge and spin degrees of freedom contribute, and decreases progressively in the I regime, as spin contributions become dominant [3]. In the $\rho\to 1$, $U>>2\Delta$ limit, the model reduces to the Heisenberg antiferromagnet with a spin-Peierls instability. The N phase is conditionally stable against dimerization.

At the NIT, the valence transition of the rigid lattice is connected in complex ways to the dimerization transitions of linear chains according to Eqn. (1). Here we present results of diagrammatic valence bond calculations [4] performed on 10-site systems with periodic boundary conditions (PBC). To reduce the dimension of the basis we impose $U, \Delta \to \infty$, with finite $\Gamma = \Delta - U/2$, in such a way to exclude doubly-ionized sites [2]. Fig. 1 shows the $\Gamma$-dependence of the equilibrium ionicity and dimerization amplitude, calculated for systems with different Peierls coupling $\varepsilon_d$ and small Holstein coupling ($\varepsilon_{sp}= 0.28$) that ensures a continuous $\rho$ variation. For $\varepsilon_d=0.64$ dimerization sets in far in the N regime, at $\rho \sim 0.25$, and the maximum amplitude $\sim 0.4$ is attained at intermediate ionicity. For smaller $\varepsilon_d=0.21$, dimerization occurs in close proximity to the NIT and its amplitude is reduced. Dimerization affects the evolution of $\rho$: with increasing $\varepsilon_d$, the $\rho(\Gamma)$ curve becomes smoother, and for $\varepsilon_d=0.64$ there is no hint in $\rho(\Gamma)$ of an N to I transition.

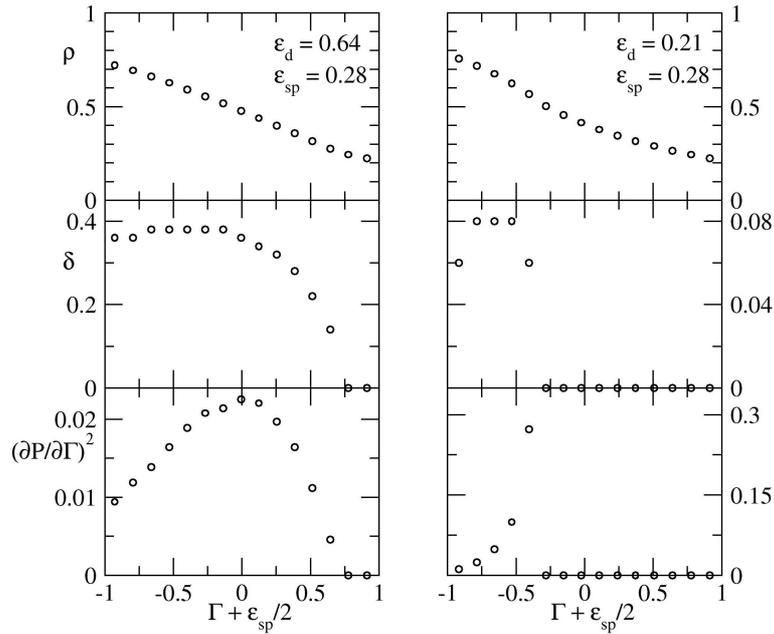

Fig. 1. The equilibrium ionicity, dimerization amplitude, and the squared derivative of the GS polarization on $\Gamma$, for a 10-site system with weak Holstein coupling and variable Peierls coupling. Energies in $t=1$ units. Note the different scales in the lower two panels.

Holstein modes, corresponding to totally symmetric vibrations of either the D or A molecule, are not infrared (IR) active in regular stacks due to inversion symmetry in the sites, but acquire intensity in dimerized phases. The IR intensity of coupled molecular vibrations (vibronic bands) is a convenient experimental probe of the evolution of $\delta$ in systems with NIT. But, up to now, no quantitative results on the intensity of vibronic bands were available. We exploit the recent definition of the GS polarization, $P$, for systems with PBC [5] to calculate IR intensities from its derivatives. The bottom panels in Fig. 1 reports the squared derivative of $P$ on $\Gamma$, the intensity of Holstein vibration being proportional to it through $\omega_v \varepsilon_{sp}$; a similar relation holds in the multimode case, but intensity borrowing is expected among vibrations with nearby frequencies [6].

The system in left panels in Fig. 1 shows vibronic bands appearing far in the N regime, with smoothly increasing intensity that attains a maximum value at intermediate ionicity. After dimerization, $\rho$ progressively increases reaching the I regime (formally $\rho>0.5$) without any hint of a phase transition. This situation is similar to that observed for ClMePD-DMeDCNQI by lowering temperature (T) or increasing pressure (P) [7]. For lower $\varepsilon_d$ (right panels), vibronic bands appear basically in the same region where the NIT can be located in terms of the maximum slope of the $\rho(\Gamma)$ curve. Fig. 1 demonstrates that the IR intensity of vibronic bands is not directly related to the dimerization amplitude. In fact, the IR intensity is larger the more responsive the charge degrees of freedom are with respect to vibrations, so that it strongly and non-linearly depends on model parameters. The intensity of vibronic bands for the system with larger $\varepsilon_d$ and larger $\delta$ is an order of magnitude smaller than for the case of smaller $\varepsilon_d$: dimerization localizes electronic charges, then reducing the electronic response. Similarly, the IR intensity of vibronic bands is, for each $\varepsilon_d$, more peaked at intermediate ionicity than the dimerization amplitude. This effect is more important at small $\delta$, where charge fluctuations are large.

Since molecular vibrations depend on $\rho$, they are affected by the NIT and can play an active role in determining the nature of the transition. Any $\rho$ variation leads to a readjustment of the molecular geometry, that in turn modifies the on-site energies, then affecting $\rho$. This self-consistent feedback mechanism amplifies the non-linearity of the system with an increase of the $\rho(\Gamma)$ slope, and a bistability region eventually appears: for large enough $\varepsilon_{sp}$ the NIT becomes discontinuous [3]. Fig. 2 reports the same systems as in Fig. 1, but for larger Holstein coupling. In the case of $\varepsilon_d=0.21$ (right panels of Fig. 2), the basic effect of increasing $\varepsilon_{sp}$ is to exclude a region of intermediate ionicity. In this case one observes a discontinuous $\rho$ variation, with vibronic bands appearing abruptly as the system crosses the interface. A similar behaviour is observed for T or P induced phase transitions of TTF-CA [8]. The case with for $\varepsilon_d=0.64$ (left panels in Fig. 2) is different. Although the $\rho(\Gamma)$ variation is stronger than in Fig. 1, even a large Holstein coupling ($\varepsilon_{sp}=2.8$) does not produce a discontinuous transition. It turns out to be difficult, if at all possible, to find system parameters leading two separate transitions: a dimerization transition in the N phase and subsequently a discontinuous NIT [9]. No such behaviour has been observed so far. A double transition has been reported for TTF-CA in low-T and high-P experiments [10], but finite-T effects have to be considered to properly address this phenomenon.

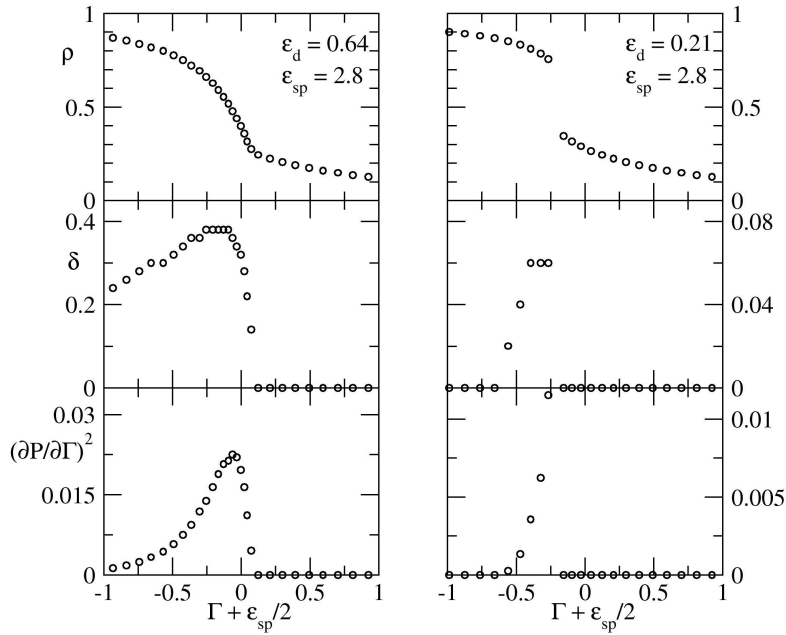

Fig. 2. The same as Fig. 1, but for stronger Holstein coupling.

We have calculated the evolution of equilibrium ionicity and dimerization amplitude for systems at NIT and have reported the first calculation of IR intensity of vibronic bands. Our results demonstrate that both Holstein and Peierls coupling are important in defining the nature of NIT and the overall properties of CT salts. Moreover the appearance of vibronic bands in IR spectra is a reliable indicator of stack dimerization, but the intensity of these bands is by no means linear with $\delta$: it in fact measures the charge displacement induced by vibrations and is strongly dependent on the electronic system. In this paper we have not included e-e interactions along the stack, nor 3D Madelung-type interactions. These interactions are important and are known to add to Holstein coupling to drive the system towards discontinuous NIT [3]. However their effect on the responses of the electronic system, and then on several crucial properties of the materials are different from those related to phonons, basically in view of the different time-scales of the two interactions. An extensive analysis of electrostatic contributions is in order to model CT salts more accurately.

Work in Parma supported by Italian MURST and CNR; work in Princeton supported by NSF MRSEC.